\documentclass[aps,prl,10pt,twocolumn,showpacs,preprintnumbers,amsmath,amssymb,superscriptaddress]{revtex4-1} 

\usepackage{graphicx}
\usepackage{dcolumn}
\usepackage{color}
\usepackage{amsmath}
\usepackage[breaklinks,colorlinks,bookmarks=false,citecolor=blue,linkcolor=red,urlcolor=blue]{hyperref}

\newcommand{\bibtitle}[1]{{\slshape #1},} % Title ON

\renewcommand{\vec}[1]{{\mathbf #1}} 

\renewcommand{\onlinecite}[1]{\cite{#1}}

\begin{document}

\preprint{\href{https://doi.org/10.1103/PhysRevB.96.060403}{Phys. Rev. B {\bfseries 96}, 060403(R) (2017)}}

\title{Finite-Temperature Dynamics and Thermal Intraband Magnon Scattering\\
in Haldane Spin-One Chains}

\author{J. Becker}
\affiliation{Institut f\"ur Theoretische Festk\"orperphysik, JARA-FIT and JARA-HPC, RWTH Aachen University, 52056 Aachen, Germany}

\author{T. K\"ohler}
\affiliation{Institut f\"ur Theoretische Physik, Georg-August-Universit\"at G\"ottingen, 37077 G\"ottingen, Germany}

\author{A.C. Tiegel}
\affiliation{Institut f\"ur Theoretische Physik, Georg-August-Universit\"at G\"ottingen, 37077 G\"ottingen, Germany}

\author{S.R. Manmana}
\affiliation{Institut f\"ur Theoretische Physik, Georg-August-Universit\"at G\"ottingen, 37077 G\"ottingen, Germany}

\author{S. Wessel}
\affiliation{Institut f\"ur Theoretische Festk\"orperphysik, JARA-FIT and JARA-HPC, RWTH Aachen University, 52056 Aachen, Germany}

\author{A. Honecker}
\affiliation{Laboratoire de Physique Th\'eorique et Mod\'elisation, CNRS UMR 8089, Universit\'e de Cergy-Pontoise, 95302 Cergy-Pontoise Cedex, France}

\date{March 23, 2017; revised May 29, 2017} %\today}

\begin{abstract}
The antiferromagnetic spin-one chain is considerably one of the most 
fundamental quantum many-body systems, with symmetry protected topological 
order in the ground state. Here, we present results for its dynamical spin 
structure factor at finite temperatures, based on a combination of exact 
numerical diagonalization, matrix-product-state calculations and quantum 
Monte Carlo simulations. Open finite chains exhibit a sub-gap band in the 
thermal spectral functions, indicative of localized edge-states. Moreover, 
we observe the thermal activation of a distinct low-energy continuum 
contribution to the spin spectral function with an enhanced spectral 
weight at low momenta and its upper threshold. This emerging thermal 
spectral feature of the Haldane spin-one chain is shown to result from 
intra-band magnon scattering due to the thermal population of the 
single-magnon branch, which features a large bandwidth-to-gap ratio. These 
findings are discussed with respect to possible future studies on spin-one 
chain compounds based on inelastic neutron scattering.
\end{abstract}


\maketitle

%\section{Introduction}
%\label{sec:intro}

One-dimensional quantum spin models constitute basic condensed matter 
many-body systems that despite their simplicity exhibit a rich variety of 
emergent phenomena~\cite{Mikeska04}. These include the formation of 
collective excitations and non-classical ground states with characteristic 
patterns in the quantum entanglement. From this perspective, Haldane's 
conjecture~\cite{Haldane81,Haldane83a,Haldane83b} on a fundamental 
difference in the low-energy physics of integer-valued spin chains with 
respect to the spin-half Heisenberg chain has established the spin-one 
chain model as a fundamental spin system, which furthermore finds 
realizations in various, mainly Ni${}^{2+}$-based compounds 
\cite{Buyers86, Steiner87, Renard87, Renard88, Regnault94, Orendac95, 
Takigawa96, Xu96, Honda98, Zheludev01prb, Xu07, Niazi09, Pieper09}. Its 
properties have been intensively explored in both theoretical and numerical, 
as well as experimental studies in recent years, mainly with a focus 
toward the peculiar properties of the gapped ground 
state~\cite{Nijs89,Kennedy92}, which is now understood as a most basic 
instance of symmetry protected topological (SPT) 
order~\cite{Gu09,Pollmann10}. This leads, e.g., to the formation of a pair 
of entangled spin-half low-energy edge states for open finite 
chains~\cite{Kennedy90}.

\begin{figure}[t]
	\includegraphics[width=0.45\textwidth]{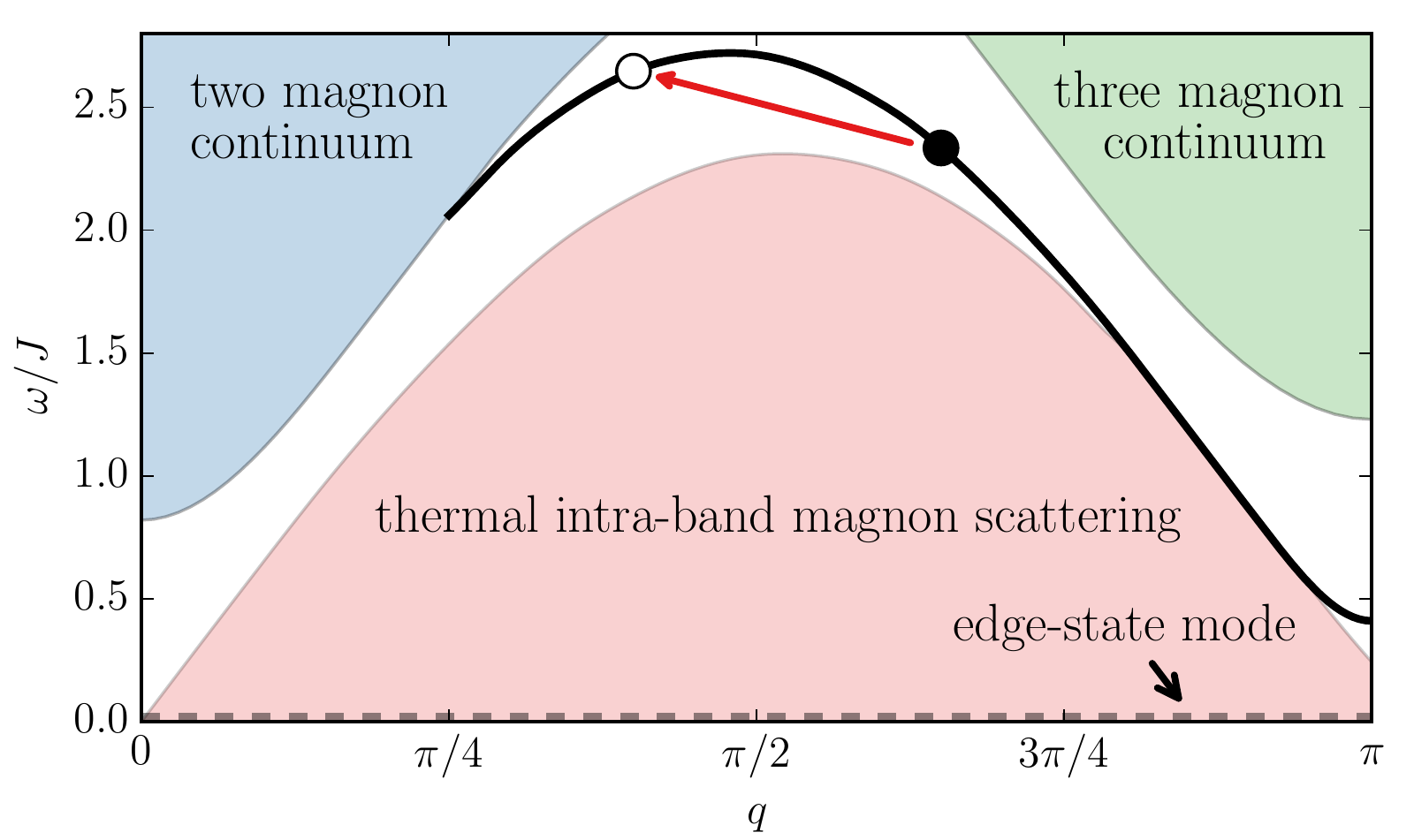}
	\caption{Sketch of the low-energy excitations of the Haldane spin-one chain. 
	The black line shows the gapped single-magnon dispersion, and the upper shaded regions denote the two- and three-magnon continua. The lower shaded region encloses the intra-band-magnon-scattering contribution to the dynamical spin structure factor that emerges from the thermal population of the single-magnon branch. 
	For open chains, an additional sub-gap edge-state mode extends from $q=\pi$ towards smaller momenta, indicated by the dashed line.
	}
	\label{fig:spin1_sketch}
\end{figure}

Dynamical probes of quantum magnetism in spin-one chain compounds, 
performed using inelastic neutron scattering, have confirmed the gapped 
magnetic excitation spectrum~\cite{Steiner87, Ma92, 
Cowley01,Zaliznyak01,Kenzelmann02,Xu07,Bera13}. At low temperatures, the 
corresponding dynamical spin structure factor is dominated by the gapped 
single-magnon branch, with additional contributions from multi-magnon 
continuum states, leading to the termination of the single-magnon branch 
due to decay and scattering with the two-magnon continuum states 
\cite{Nightingale86,Takahashi89,GomezSantos89,SaTa90,Deisz90,Affleck92,White92,White93a,White93b, 
Meshkov93,Golinelli94,Yamamoto97,Schmitt98,Horton99,Essler00,Kenzelmann01b,Kolezhuk06,White08,Rahnavard15}, 
cf.\ Fig.~\ref{fig:spin1_sketch} for an illustration. The effects of 
thermal fluctuations on the dynamical spin structure factor at elevated 
temperatures~\cite{Kenzelmann01a,Kenzelmann02b,Xu07} have been less 
intensively investigated theoretically, in particular in the region of 
intermediate energy scales, where theoretical approaches require one to 
account for both quantum and thermal fluctuations. Previous theoretical 
works mainly focused on the temperature-induced shift in the single-magnon 
dispersion as well as its thermal broadening in the low-temperature 
regime~\cite{Jolicoeur94,Damle98,Kenzelmann02a,Syljuasen08,Essler08,Essler09}.

In this Rapid Communication, we discuss the emergence of a distinct, thermal contribution to the finite-temperature dynamical spin structure factor that we find to result from intra-band magnon scattering (IBMS), cf.\ Fig.~\ref{fig:spin1_sketch}.
The IBMS continuum exhibits an 
enhanced spectral weight near its upper edge, resulting 
from the van-Hove singularity in the density of states near the extrema
of the single-magnon band.
This enhanced spectral weight appears close to the single-magnon branch
due to the large bandwidth of the latter.
Our results furthermore indicate that this 
 thermal IBMS may feasibly be detected upon performing neutron scattering experiments in a temperature regime of the order of the spin gap. 
In addition, we find a signature of an edge-state mode for open chains, which is visible over an extended temperature region. 

Before presenting our results for the dynamical spin structure factor, we first introduce the model and the employed numerical methods. 
The Hamiltonian for the SU(2)-symmetric antiferromagnetic spin-one chain of length $L$ reads
$ %\begin{align}
H = J \sum_{\langle{} i,j \rangle{}} \vec{S}_i \cdot \vec{S}_j , 
$ %\label{eq:hamiltonian}
%\end{align}
with $J>0$,
where in the following we employ both open chains (OBC) and closed chains with periodic boundary conditions (PBC). The 
dynamical spin structure factor is given in the Heisenberg picture as 
%\begin{align}
$
S(q,\omega) = \int\mathrm{d}t \, \mathrm{e}^{-\mathrm{i}\omega t} 
 \left\langle \vec{S}_{q}(t) \cdot \vec{S}_{-q}(0) \right\rangle,
$
%\label{eq:structurefactor}
%\end{align}
where $\vec{S}_q=\frac{1}{\sqrt{L}} \sum_j \mathrm{e}^{-\mathrm{i} q j } \vec{S}_j$, with $q=2\pi\nu/L$, $\nu=1,2,...,L$ for PBC.
Using numerical exact diagonalization (ED), we were able to obtain numerically exact results for $S(q,\omega)$ on finite chains with PBC up to $L=20$~\cite{Honecker16,SM,Lanczos50,RevModPhys.66.763,JPC72,PhysRevLett.59.2999}. 
In order to access larger system sizes, we used both density-matrix renormalization group (DMRG) \cite{White92,White93a,Schollwoeck05} and quantum Monte Carlo (QMC) \cite{Sandvik10} approaches to calculate $S(q,\omega)$. 
For the DMRG-based ana\-lysis we used a recently developed finite-temperature approach \cite{Tiegel14}, formulated within matrix product states (MPS) \cite{Schollwoeck11}, which works directly in the frequency domain.
As is the case for other finite-temperature time-dependent DMRG algorithms \cite{Verstraete04,Feiguin05,Barthel09},
this method is based on the purification of the thermal density operator obtained via imaginary time evolution. 
However, the underlying thermofield formalism \cite{Barnett87} in combination with Liouville-space dynamics \cite{Dalton82} allows us to naturally work in the frequency domain 
and thus apply a moment expansion in terms of Chebyshev polynomials to the spectral function itself \cite{Weisse06, Holzner11, Schmitteckert14}.
Working with OBC in the DMRG calculations for efficiency reasons, the momentum-space spin-operators are related to those in real space via
$
\vec{S}_q = \sqrt{\frac{2}{L+1}} \sum_{j=1}^L \sin\left( q j \right) \vec{S}_j,
$
where $q=\pi\nu/(L+1)$, $\nu=1,2,...,L$~\cite{Benthien04}.
We typically consider a chain length of $L=32$ and
an MPS truncation at bond dimension $m=250$ 
which yields compression errors $\mathcal{O} (10^{-2})$.
The iterative Chebyshev expansion is truncated at order $2000$, which results in an estimated broadening $\sigma_\omega$, weakly frequency dependent, of the order of $0.1J$.
For the QMC calculations 
we used the stochastic series expansion (SSE) algorithm with a generalized directed loop update~\cite{Sandvik02,Alet05}, and both OBC and PBC can be considered equally well.
In order to access the spin dynamics, correlation functions
in Matsubara frequency space,
$
	C(q,\mathrm{i}\omega_n) = \int_{0}^{\beta} \mathrm{d}\tau\, 
	\mathrm{e}^{\mathrm{i}\omega_n\tau} 
	\left\langle \vec{S}_q(\tau) \cdot \vec{S}_{-q}(0)\right\rangle,
	\label{eq:correlationfunction}
$
with 
$\omega_n=2\pi n/\beta$, $n\in \mathbb{N}_0$
are measured, utilizing a mapping of the SSE configuration-space to continuous imaginary time~\cite{Sandvik97,Michel07}.
Here, $\beta=1/T$, and we typically require up to the 200 lowest Matsubara frequencies. 
Real-frequency spectra are then obtained by performing an analytic continuation 
to invert the relation
$
C(q,\mathrm{i}\omega_n) = \int_0^\infty \mathrm{d}\omega\,\frac{\omega}{\pi}
\frac{1-\mathrm{e}^{-\beta\omega}}{\omega_n^2+\omega^2} S(q,\omega).
$
To this end, we employ a stochastic analytic continuation algorithm \cite{Beach04} which yields Monte Carlo averages over ensembles of proposed spectral functions.

\begin{figure}[t]
\includegraphics[width=0.5\textwidth]{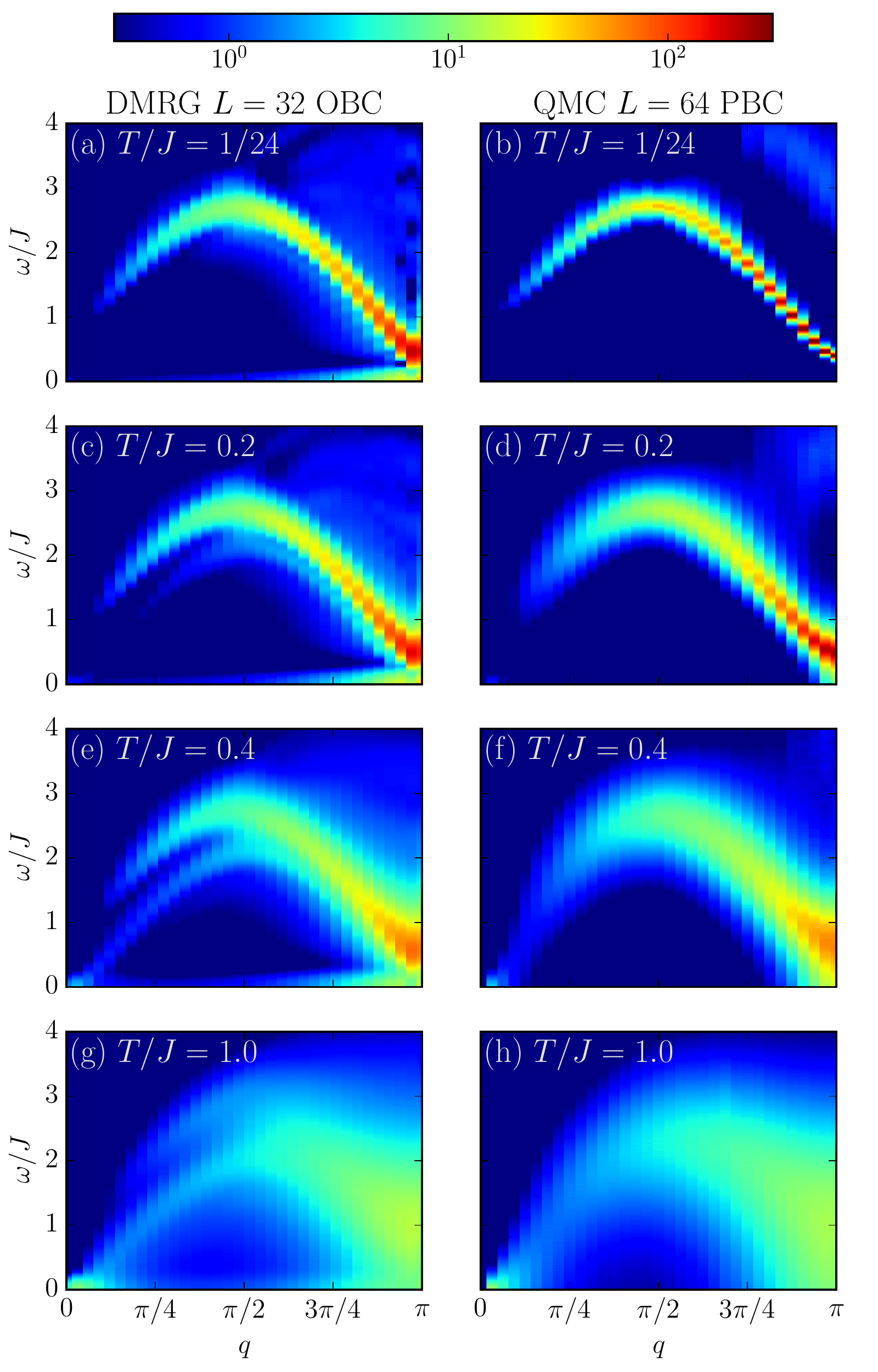}
\caption{Dynamical spin structure factor $S(q,\omega)$ for the Haldane spin-one chain from DMRG with OBC (left panels), and QMC with PBC (right panels) for various temperatures $T$. }

 \label{fig:specs_dmrg_qmc} 
\end{figure}

An overview of our main findings, the spectral function $S(q,\omega)$ of the spin-one chain at different temperatures, is provided in 
Fig.~\ref{fig:specs_dmrg_qmc}, where the left (right) column shows DMRG (QMC) results for a chain with OBC (PBC).
A comparison of the DMRG spectral functions at a set of fixed momenta and for different temperatures is also available~\cite{SM}.
 The data obtained by our finite-temperature schemes at 
$T/J=1/24$ (panels (a) and (b)) effectively represents ground state results. The most prominent contribution to $S(q,\omega)$ is the single-magnon branch, with a lowest excitation gap of $\Delta\approx{}0.41J$ at the antiferromagnetic wave vector, $q=\pi$
\cite{Nightingale86,White93b,Golinelli94}. Near $q=\pi/4$, the magnon branch
merges into the two-magnon continuum, leading to the decay of elementary magnon
excitations~\cite{White93b,Kolezhuk06,White08}. Correspondingly, in the low-$q$
region, we observe a loss of spectral weight.
For a finite system with OBC (cf.\ Fig.~\ref{fig:specs_dmrg_qmc} (a)), a distinct additional contribution to the spin dynamics results from the low-energy edge states located at the two ends of an open spin-one chain~\cite{Kennedy90}. 
Due to the local character of the edge-state contribution, this low-energy spectral weight vanishes proportional to $1/L$ upon increasing the system size. This is confirmed by a finite-size analysis of the total spectral weight in the sub-gap region~\cite{SM}.
In calculations with PBC, this sub-gap feature is absent (cf.\ Fig.~\ref{fig:specs_dmrg_qmc} (b)), while for chains with OBC we also obtain it from QMC~\cite{SM}.
The DMRG spectral function in Fig.~\ref{fig:specs_dmrg_qmc} (a) shows a tiny fraction of the spectral weight which is spread both below and above the single-magnon branch. 
This results mainly from the truncation of the Chebyshev expansion
and the comparatively small MPS bond dimension, and
is not observed in the QMC simulations. The QMC spectrum in Fig.~\ref{fig:specs_dmrg_qmc} (b) thus allows us to also resolve the well-separated three-magnon continuum near $q=\pi$, where its intensity is sufficiently large~\cite{SM}.

\begin{figure}[!t]
\includegraphics[width=0.5\textwidth]{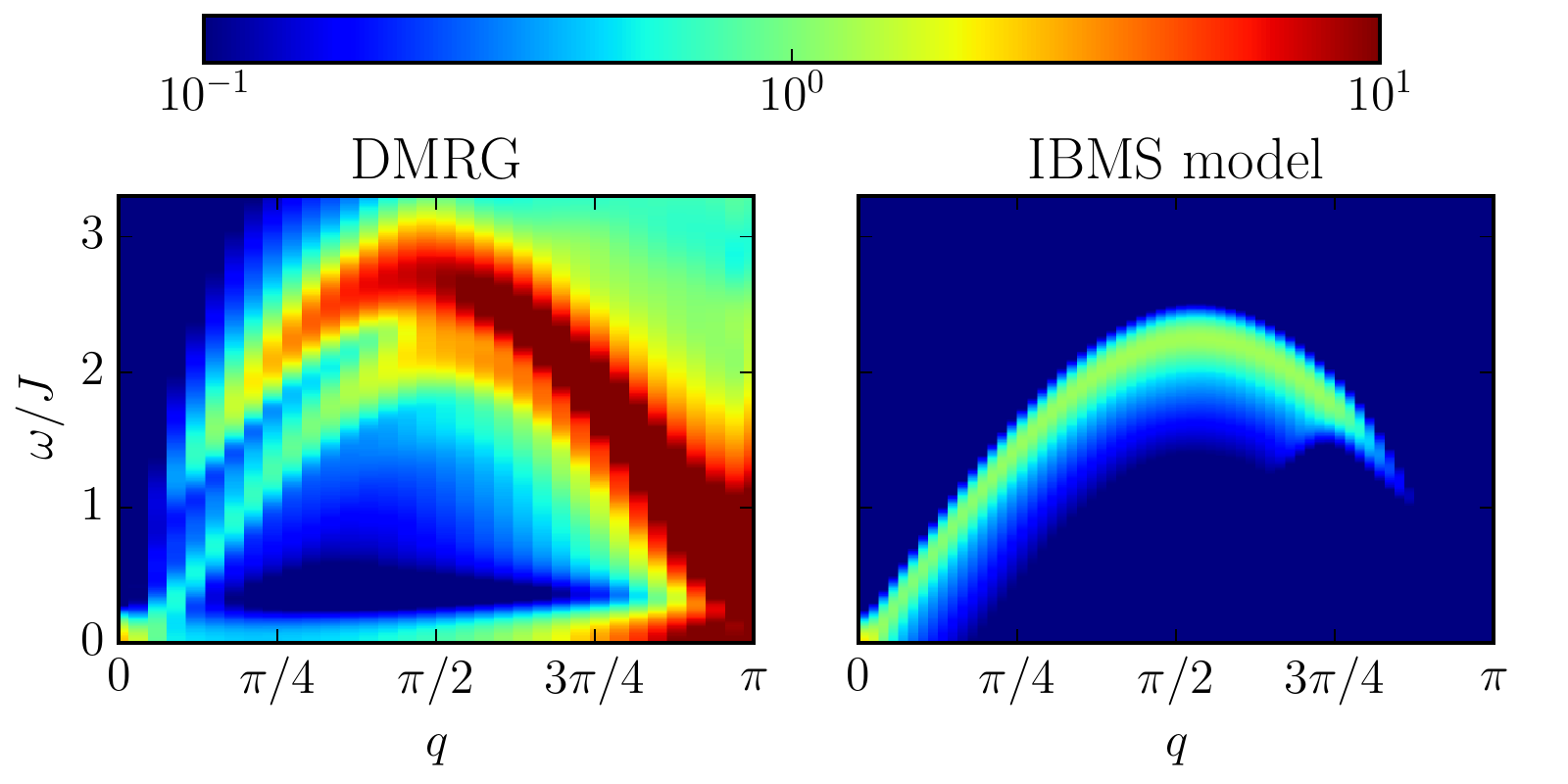}
\caption{
Comparison of the DMRG spectral function $S(q,\omega)$ (left panel) for $L=32$ (OBC) with $S^{IB}(q,\omega)$ calculated for the IBMS model (right panel) at $T/J=0.3$. 
A Gaussian broadening with $\sigma_\omega=0.1J$, similar to the DMRG spectra, was applied to the IBMS model spectral function.}
 \label{fig:ib_spec_model_dmrg}
\end{figure}

We next consider the thermal effects on the dynamical spin structure factor,
cf.\ Fig.~\ref{fig:specs_dmrg_qmc} (c)--(h), as well as Fig.~\ref{fig:ib_spec_model_dmrg}. The thermal broadening of the single-magnon branch as well as the thermal band narrowing has been examined previously~\cite{Jolicoeur94, Damle98,Syljuasen08}, cf. also Ref.~\onlinecite{SM}. 
The OBC spectra furthermore show that the open finite-system's edge-state contribution to the dynamical spin structure factor remains a distinct sub-gap feature also at finite temperatures, which thus provides a convenient fingerprint of the SPT nature of the ground state. 

A qualitative change seen only in the finite-$T$ spectral function is the emergence of additional spectral weight below the single-magnon branch for $T\gtrsim \Delta/2\approx 0.2J$, which is well separated from the single-magnon branch for $q\lesssim\pi/2$. At $T=0.4J$, cf.\ Fig.~\ref{fig:specs_dmrg_qmc} (e), this temperature-induced spectral weight still appears to resemble a dispersing mode, softening at $q=0$, where the spectral weight is further enhanced. 
While the DMRG approach allows us to distinguish this temperature-induced spectral weight from the single-magnon branch, the spectral function obtained from the analytically continued QMC data (cf.\ Fig.~\ref{fig:specs_dmrg_qmc} (f))
is affected by a difficulty of the analytic continuation to separate such closely spaced spectral weight contributions at finite temperatures. 
The QMC data nevertheless exhibit
the presence of the thermal spectral weight contribution
at low energies, close to $q=0$. 
Upon further increasing the temperature, a redistribution of the spectral weight
can be seen in Fig.~\ref{fig:specs_dmrg_qmc}, and this eventually reveals the actual character of the temperature-induced spectral feature, which forms an extended continuum with an enhanced spectral weight at its upper threshold (cf.\ Fig.~\ref{fig:specs_dmrg_qmc} (g) and (h)).

This thermal spectral weight results from IBMS processes that have been previously
observed in dimerized spin-1/2 chains \cite{Notbohm07,James08,Tennant12}: the thermal population of the magnon mode, predominantly in the vicinity of $q=\pi$, where the magnon dispersion has its lowest excitation gap, allows for scattering processes of a thermally excited magnon to another state on the single-magnon branch, 
cf.\ the illustration in Fig.~\ref{fig:spin1_sketch}.
Such processes contribute to $S(q,\omega)$ upon respecting the conservation of
momentum and energy exchange with the scattering particle (such as, e.g., in neutron scattering).
More quantitatively, 
this thermal IBMS contribution $S^{IB}(q,\omega)$ to the dynamical spin structure factor 
can be approximately obtained using a magnon-state representation within a basic kinematic model.
We denote by $| k ,\sigma\rangle $ a single-magnon ($S_\mathrm{tot}=1$) excitation of momentum $k$ and $S^z_\mathrm{tot}=\sigma\in\{ 0,\pm 1\}$ atop the $S_\mathrm{tot}=0$ ground state $|0\rangle$, with an excitation energy $\epsilon_k$ along the single-magnon branch. 
The multi-magnon states are subject to a hard-core constraint that can
be treated in several approximate ways that all yield the
same low-temperature asymptotics.
%To treat IBMS in the regime $T\lesssim \Delta$,
We found it convenient to use a $k$-space-based hard-core boson approximation
of the initial $(i)$ and final $(f)$ states 
in the Lehmann representation of 
$
S(q,\omega)=3\sum_{i,f} \mathrm{e}^{-\beta E_i}/Z\: |\langle f |S^z_q | i\rangle |^2 \delta(\omega-E_f+E_i).
$
Here, the factor of three accounts for the SU(2) symmetry of the Hamiltonian $H$. 
Neglecting further interaction effects, $E_i$ ($E_f$) equals the sum
of the occupied single-magnon state energies in the initial (final)
state, and the partition function
$Z=\prod_{k,\sigma} (1+ \mathrm{e}^{-\beta\epsilon_k})$. 
The leading-order scattering processes, whereby a thermally excited magnon
is scattered into another unoccupied single-magnon state, then yield 
%
%\begin{equation}
\[
S^{IB}(q,\omega)=3\sum_{k,\sigma} \frac{ |\langle k+q,\sigma | S^z_q | k,\sigma \rangle|^2 }{(1+ \mathrm{e}^{\beta\epsilon_k})(1+ \mathrm{e}^{-\beta\epsilon_{k+q}})} \delta(\omega -\epsilon_{k+q}+\epsilon_k).\label{eq:sib}
\]
%\end{equation}
%
Finally, we approximate the nonvanishing scattering matrix elements as 
$
|\langle k+q, \pm 1 | S^z_q | k, \pm 1 \rangle|^2 \approx 1/L,
$
which would hold exactly, if the single-magnon states were obtained as 
$|k,\pm 1 \rangle=S^\pm_k |0\rangle$ and $ S^z_q|k,0\rangle=0$,
using that 
$[S^z_q,S^\pm_k]=\pm S^\pm_{k+q}$, 
with
$S^\pm_q=\frac{1}{\sqrt{L}} \sum_j \mathrm{e}^{-\mathrm{i} q j } S_j^\pm$.
The overall $1/L$-scaling of the matrix elements renders $S^{IB}(q,\omega)$ convergent in the thermodynamic limit. 
In addition to the above explicit treatment of the longitudinal ($S_q^z$) channel,
one can also perform a similar calculation for the 
transverse sectors of $S^{IB}(q,\omega)$, which then indeed exhibits its anticipated SU(2) symmetry.

\begin{figure}[!t]
\includegraphics[width=0.5\textwidth]{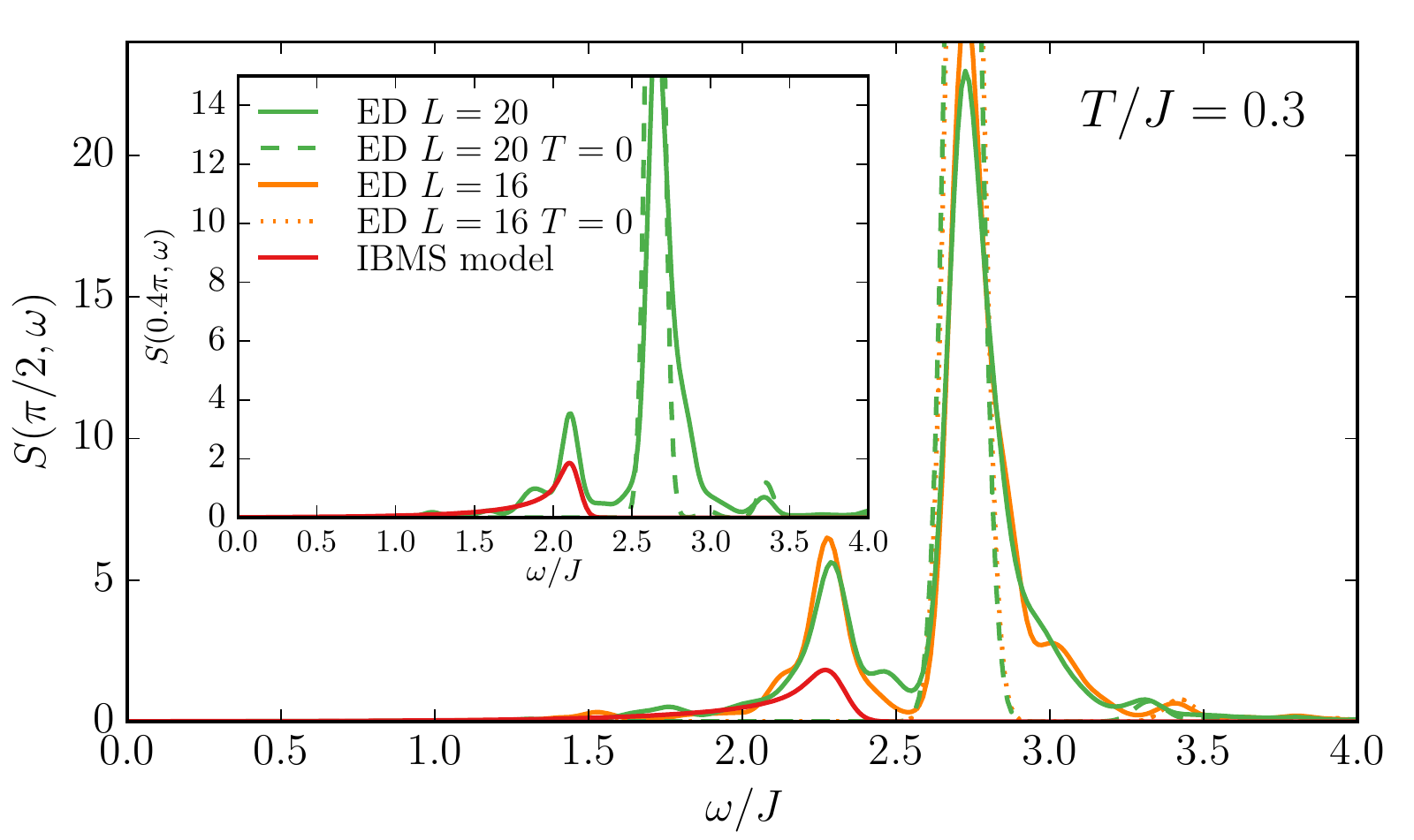}
\caption{
Comparison of the ED spectral functions $S(q,\omega)$ with the IBMS model for $S^{IB}(q,\omega)$ at $T/J=0.3$.
The main panel shows results for $q=\pi/2$, and the inset those for $q=0.4\pi$. 
For comparison, ED results for $T=0$ are also included. 
A Gaussian broadening with $\sigma_\omega=0.05J$ was applied to all spectral functions in this figure. }
 \label{fig:ed_ibms}
\end{figure}

We evaluated the IBMS contribution from this basic model, based on the single-magnon dispersion taken from Ref.~\onlinecite{White08}. 
The resulting IBMS spectral function at $T/J=0.3$ is shown in the right panel of Fig.~\ref{fig:ib_spec_model_dmrg}, next to the corresponding DMRG result for $S(q,\omega)$. 
Here, we convoluted the IBMS model spectral function with a Gaussian resolution of width $\sigma_\omega=0.1J$, i.e., the broadening in the DMRG spectral functions.
We find that our rather simple model qualitatively captures the shape of the IBMS contribution, in particular its upper boundary. 
Near this threshold, as well as near $q=0$, the spectral weight is enhanced due to the van-Hove singularity in the magnon density of states near $k=\pi/2$ and $\pi$. 
The full extent of the IBMS continuum as obtained within the IBMS model is indicated in Fig.~\ref{fig:spin1_sketch}.
Within the maximum energy regime $\omega/J\approx 2$ of the IBMS signal near $q=\pi/2$, where finite-size effects are expected to be weakest, we can use the $L=20$ ED data for a more detailed comparison, 
since in the ED approach, we can choose a smaller broadening $\sigma_\omega=0.05J$. 
A comparison of the ED spectral functions for $q=\pi/2$ and $q=0.4\pi$ to the IBMS model is shown in Fig.~\ref{fig:ed_ibms} for $T/J=0.3$. 

For $q=\pi/2$, where we can directly compare ED data for $L=20$ and $L=16$ (since for both chain lengths, $q=\pi/2$ is an available lattice momentum) we conclude that indeed the $L=20$ data in the relevant energy region 
exhibit only weak residual finite-sizes effects. By a direct comparison to the $T=0$ data, we identify the thermally induced spectral weight, with a peak at $\omega/J\approx 2.3$, and clearly separated from the magnon peak at $\omega/J\approx 2.7$.
The position of the thermal peak is well reproduced by the IBMS model. To compare the corresponding spectral weight in the ED data to the IBMS model, one needs to account for the 
additional weight in the ED spectral function that is due to the broadened magnon peak; this elevates the IBMS signal in the ED data as compared to the background-free IBMS model. A similar comparison for $q=0.4\pi$, a momentum that is accessible on the $L=20$ chain, is shown in the inset of Fig.~\ref{fig:ed_ibms}. Also
here, we observe that the IBMS contribution to the ED spectral function is well reproduced by the IBMS model. 
While the above basic kinematic model already captures the overall properties of the IBMS contribution to $S(q,\omega)$, it would nevertheless be interesting to account for direct magnon-magnon interactions. As mentioned above, these lead to band-narrowing and broadening of the single-magnon mode at finite temperatures and should be accounted for in a more thorough analytical description of the IBMS process. 
Furthermore, our approximate treatment of the scattering matrix elements renders the $\omega$-integrated IBMS spectral weight less $q$-dependent than observed in the numerical results,
which show an overall increase in the IBMS signal for increasing finite values of $q$ (cf.\ Figs.~\ref{fig:specs_dmrg_qmc} and~\ref{fig:ib_spec_model_dmrg}). 
Nevertheless, our basic model clearly demonstrates the mechanism behind the IBMS contribution to the dynamical spin structure factor at finite temperatures. 

Thermally activated IBMS scattering is expected to be a general 
phenomenon in gapped quantum magnets, and indeed it is known from 
dimerized spin-1/2 chains \cite{Notbohm07,James08,Tennant12}. The case of 
the Haldane spin-one chain that we have investigated in the present Rapid
Communication is characterized by a large bandwidth as compared to the gap such 
that the maximum intensity of the IBMS continuum appears close to the 
single-magnon mode. In the present case, the IBMS thus provides an 
important contribution to the finite-temperature spin dynamics at 
low-to-intermediate scattering momenta. It would be interesting to 
identify the thermal IBMS signal from the scattering intensity in 
inelastic neutron scattering experiments on spin-one chain compounds. We 
anticipate the IBMS signal to be well accessible within a temperature 
regime set by the spin excitation gap. It may however be important to 
examine the influence of a single-ion anisotropy and inter-chain 
couplings on the IBMS signal. Furthermore, we expect the reduction of the 
spin gap by an applied magnetic field to enhance the IBMS signal toward 
lower temperatures, eventually making it relevant for the 
zero-temperature longitudinal response when the Haldane gap closes.

\begin{acknowledgments}
We thank S.\ Capponi, A.E.\ Feiguin, B.\ Lake, B.\ Normand, A.W.\ Sandvik, 
and O.F.\ Sylju{\aa}sen for useful discussions. This work was supported by 
the DFG research unit FOR1807, the CRC 1073 (Project B03), and the 
Helmholtz Virtual Institute ``New states of matter and their excitations'' 
(Project No.\ VH-VI-521). We acknowledge the allocation of CPU time at JSC 
J\"ulich and RWTH Aachen University via JARA-HPC.
\end{acknowledgments}

\newpage

\clearpage
\section{Supplemental Material}
\setcounter{figure}{0}  
\renewcommand{\thefigure}{S\arabic{figure}}

\subsection{Momentum cuts of the DMRG spectral functions}
\begin{figure}[h]
\includegraphics[width=0.5\textwidth]{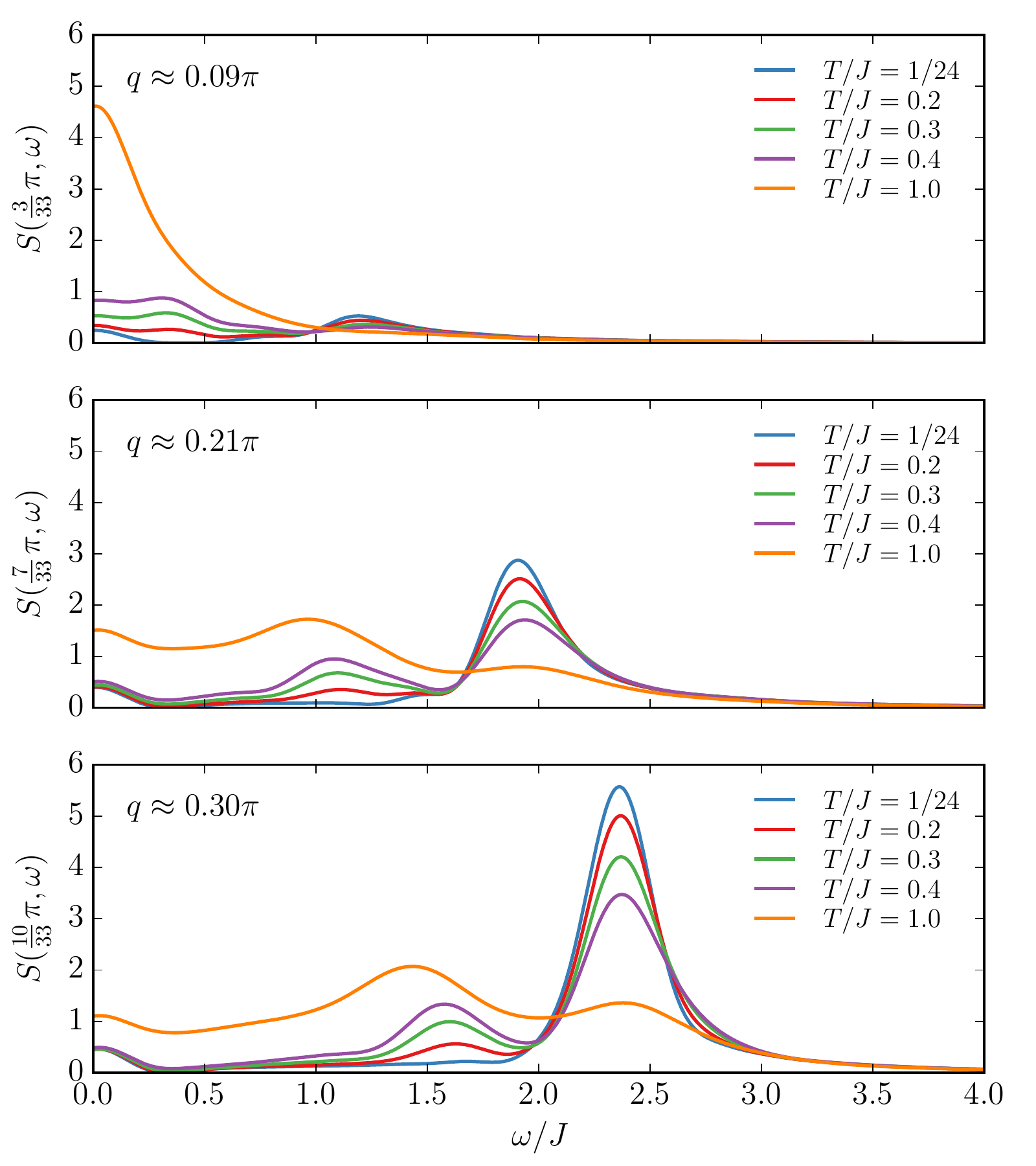}
\caption{Dynamical spin structure factor $S(q,\omega)$ from DMRG calculations for a chain length of $L=32$ with OBC for different values of $q$ and
temperatures $T/J=1/24\approx 0.0417, \ldots, 1$.
Due to the truncation of the iterative Chebyshev expansion at order $2000$, the DMRG spectra have an
estimated broadening $\sigma_\omega$ of the order of $0.1\,J$.}
 \label{fig:sDMRG} 
\end{figure}

In order to compare more directly the evolution of the dynamical spin structure factor for a given fixed momentum, we show in 
Fig.~\ref{fig:sDMRG} the spectral functions obtained from the DMRG calculations (i.e.\ the same data  as shown in
Figs.~\ref{fig:specs_dmrg_qmc} and \ref{fig:ib_spec_model_dmrg} of the main text) 
for different values of the momentum $q$ within the regime where we observe the intra-magnon scattering contribution. 

\subsection{Spectral functions from ED}

\begin{figure}[t!]
\includegraphics[width=0.5\textwidth]{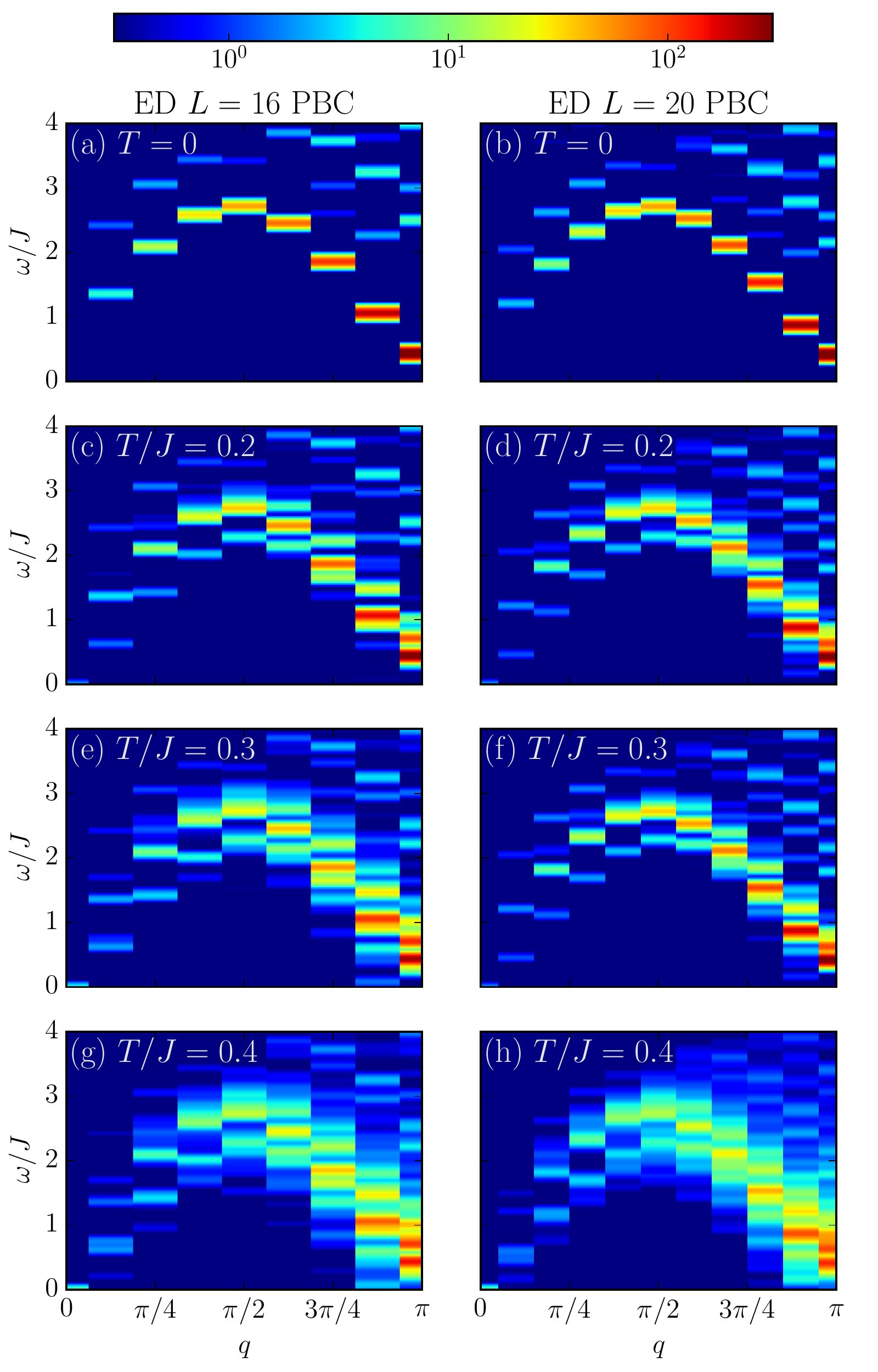}
\caption{Dynamical spin structure factor $S(q,\omega)$ from ED calculations for chain lengths of $L=16$ (left) and $L=20$ (right)  with PBC at various temperatures $T$.  A Gaussian broadening of $\sigma_\omega=0.05\,J$ was applied to the raw ED data.}
 \label{fig:sfED} 
\end{figure}

Exact diagonalization (ED) has to deal with a Hilbert space dimension
that is exponentially large in the chain length $L$.
A previous full diagonalization study \cite{Kenzelmann02b} of the spin-one chain
went up to $L=8$. We have been able to
perform full diagonalization up to $L=12$. In order to compute spectral
functions for system sizes up to $L=20$ at finite but sufficiently low
temperatures, we follow a similar strategy as in Ref.~\cite{Honecker16}.
First, we compute a certain
number of low-lying initial states $| i \rangle$ using the Lanczos
procedure \cite{Lanczos50,RevModPhys.66.763}. The thermal occupation of
each of these states is obtained with a Boltzmann weight for the
corresponding energy $E_i$. Then we apply another
Lanczos iteration to each start vector ${S}^z_q | i \rangle$
and compute the position and weight of the individual poles from the eigenvalues
and -vectors of the  tri-diagonal matrix
generated during this second Lanczos procedure \cite{RevModPhys.66.763}.
The main difference to Ref.~\cite{Honecker16} is that here we do not
employ a continued fraction expansion for the spectral function
\cite{RevModPhys.66.763,JPC72,PhysRevLett.59.2999},
but rather subject each pole to Gaussian broadening.

In Fig.~\ref{fig:sfED}, we show the ED data for the dynamical spin structure factor of the spin-one chain at different temperatures   and chain lengths with PBC. Similarly to the DMRG results shown in the main text, we can identify the IBMS contribution at finite temperatures.  

%
%\newpage
\subsection{QMC for OBC and Edge-State Contribution}

\begin{figure}[t!]
\includegraphics[width=0.5\textwidth]{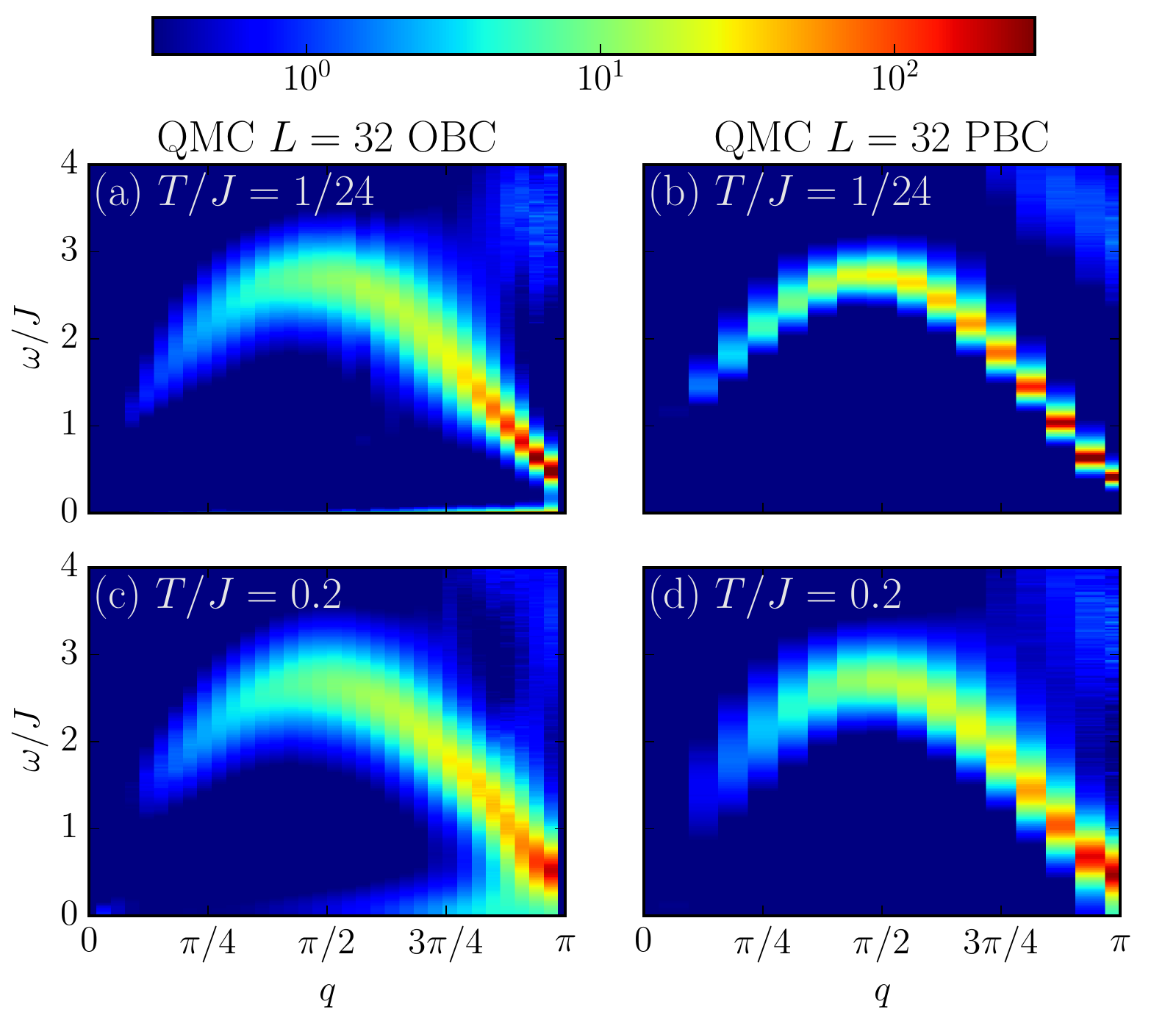}
\caption{Dynamical spin structure factor $S(q,\omega)$ from QMC
simulations with OBC as well as PBC for $L=32$ spins at two different temperatures $T$. 
}
 \label{fig:qmcobc} 
\end{figure}

\begin{figure}[t!]
\includegraphics[width=0.4\textwidth]{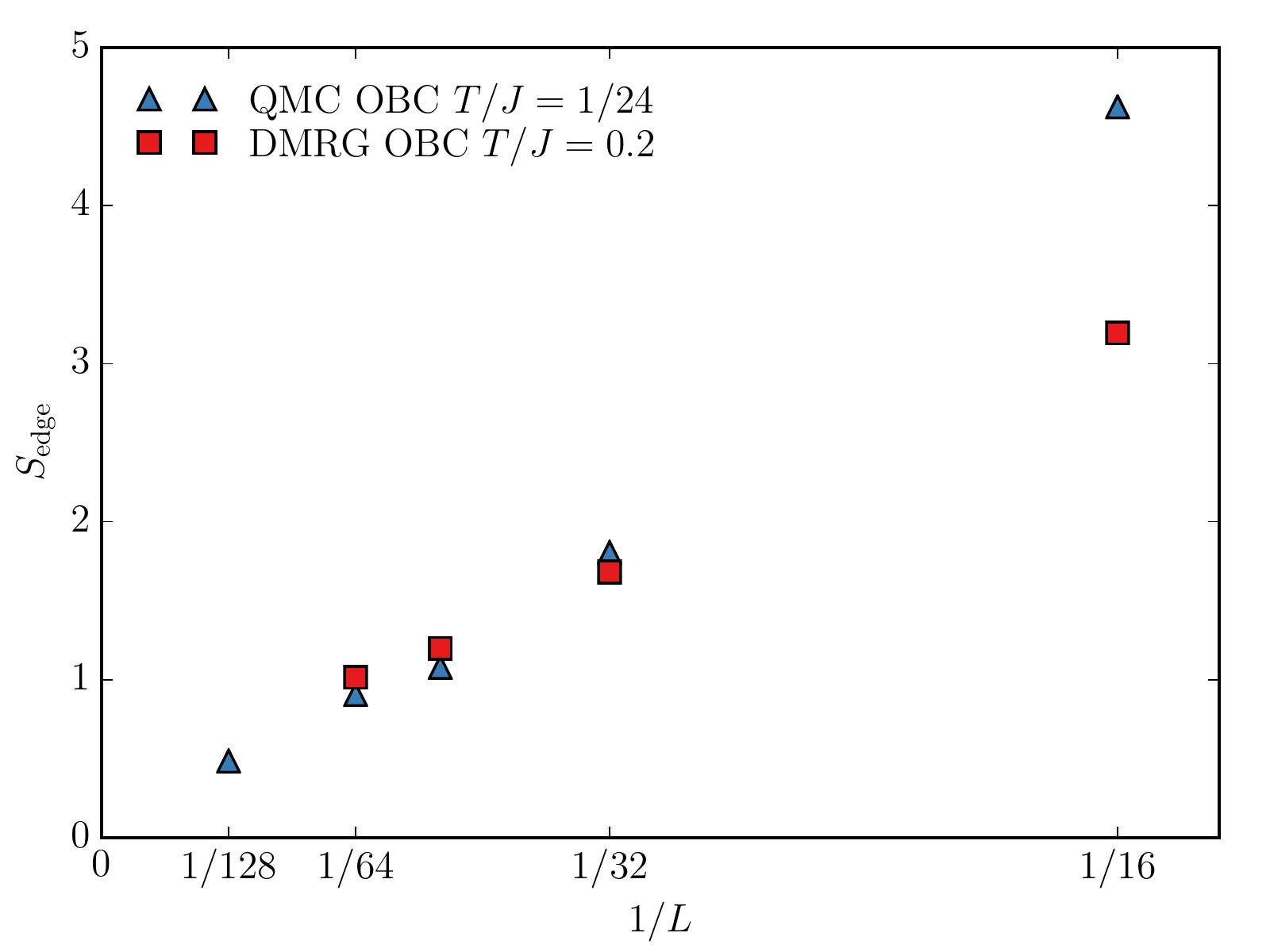}
\caption{Integrated weight of the edge-state contribution $S_\mathrm{edge}$ as a function of the inverse chain length $1/L$.}
 \label{fig:edge_mode_finite_size_scaling}
\end{figure}

In Fig.~\ref{fig:qmcobc}, we compare  the QMC results for the  dynamical spin structure factor of the spin-one chain with $L=32$ at different temperatures  for PBC and  OBC.
Like DMRG (see Fig.~\ref{fig:specs_dmrg_qmc} of the main text),
the QMC dynamical spin structure factor for OBC also 
exhibits the sub-gap mode emerging from the edge-state contribution. 
To quantify the finite-size scaling of this edge-state contribution, we performed a spectral weight integration in the sub-gap region 
to obtain the total integrated 
edge-state mode weight, calculated as $S_\mathrm{edge}=\int_0^\pi \mathrm{d}q \int_0^{\omega_{\mathrm{max}}} \mathrm{d}\omega \; S(q,\omega)
$, where $\omega_{\mathrm{max}}=0.2J$  for the QMC results and $\omega_{\mathrm{max}}=0.25J$ for the DMRG spectra,
accounting  for the  increased (thermal) broadening. The resulting finite-size scaling of $S_\mathrm{edge}$ for different temperatures and from both DMRG and QMC data is shown in Fig.~\ref{fig:edge_mode_finite_size_scaling}. This shows that the edge-state contribution vanishes as $1/L$ with the chain length, indicative of  its local character due to  the localized edge states in the OBC spin-one chain.

\begin{figure}[t!]
\includegraphics[width=0.5\textwidth]{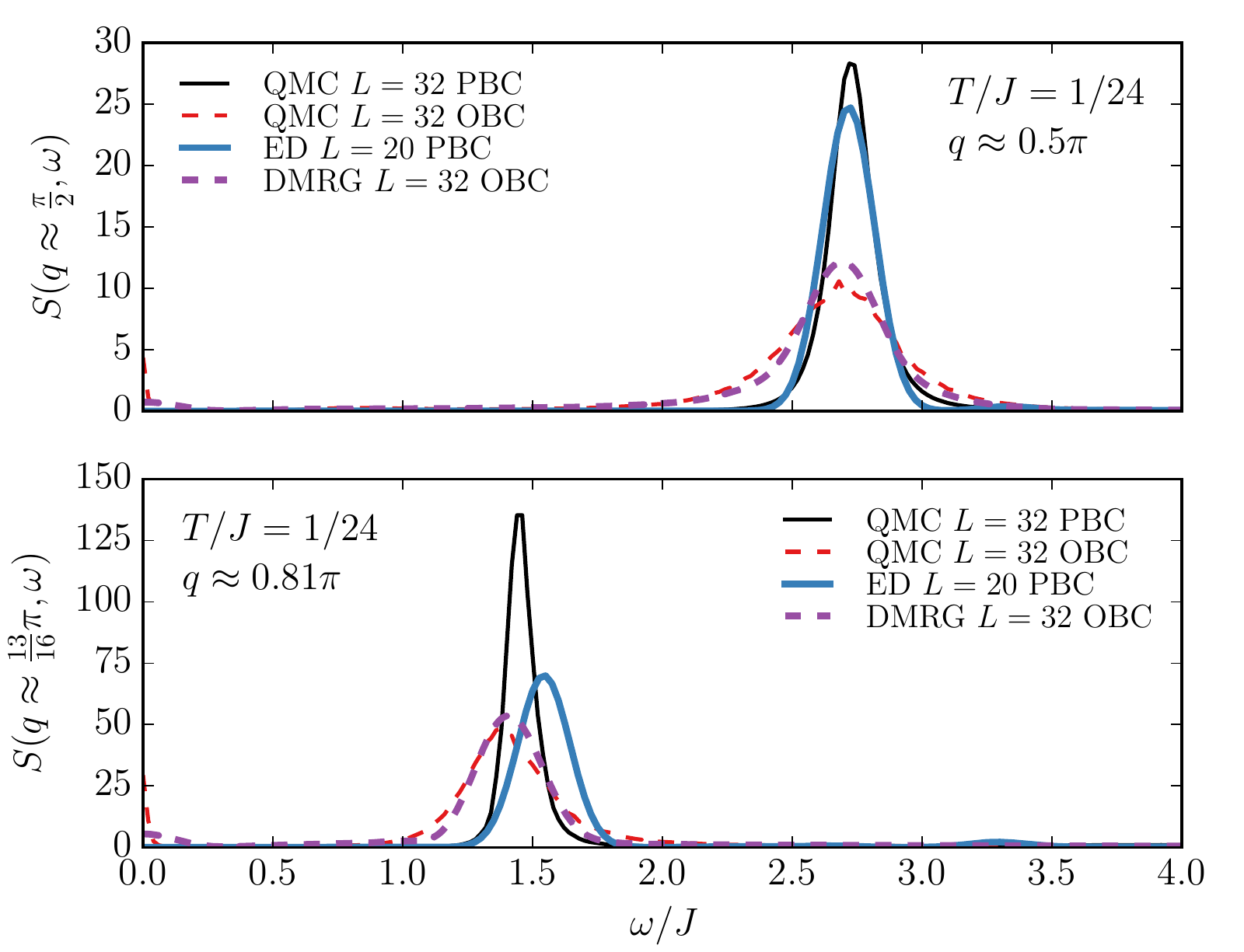}
\caption{Dynamical spin structure factor $S(q,\omega)$ 
at $T/J=1/24\approx0.0417$ from ED ($L=20$ lattice sites)
as well as QMC and DMRG calculations ($L=32$ lattice sites)
for two different momenta. ED data is subjected to a Gaussian
broadening $\sigma_\omega=0.1\,J$, the broadening of the DMRG spectra
is estimated to be of the same magnitude
due to the truncation of the Chebyshev expansion at order $2000$.}
 \label{fig:sDMRGQMC} 
\end{figure}

Figure~\ref{fig:qmcobc} also shows that the magnon line is generally broader for
OBC (panels (a) and (c)) than for PBC (panels (b) and (d)). To some extent, this may
be due to $q$ not being a good quantum number for OBC and a related mixing with neighboring
momenta, although the broadening due to mixing should be negligible at the minimum
($q = \pi$) and maximum ($q \approx \pi/2$) of the dispersion.
This broadening is investigated in more detail
in Fig.~\ref{fig:sDMRGQMC}, where we compare directly QMC, DMRG, and ED
results for the spectral functions between PBC and OBC for 
a set of  momenta $q$, with a focus on the magnon peak.
We find that  overall the spectral function, as obtained for both boundary
conditions, compare rather well between the different methods.
Just at $q \approx 0.81\,\pi$, ED yields a peak at a slightly higher $\omega$ owing
to the fact that in this case, the data is actually for $q = 0.8\,\pi$. For $q \approx 0.81\,\pi$,
QMC also evidently yields a line with a width below $\sigma_\omega=0.1\,J$ for PBC.
%The width of the magnon peak is only mildly larger for the analytically continued QMC data for OBC than from the DMRG calculation.

By comparing the OBC results to the corresponding QMC spectra for PBC
in Fig.~\ref{fig:sDMRGQMC}, we confirm that the magnon peak
has a larger width than for PBC. This  could be due to  additional scattering processes for OBC of the magnon excitations with the edge states, while for PBC, this scattering channel is
% of course 
not available.  It might  thus be interesting to analyze this effect of the edge states on the magnon excitation  in more detail in further investigations. However,  we also expect  such additional edge-state scattering  to become less relevant for larger system sizes.

%
%\clearpage

\begin{figure}[t!]
\includegraphics[width=0.5\textwidth]{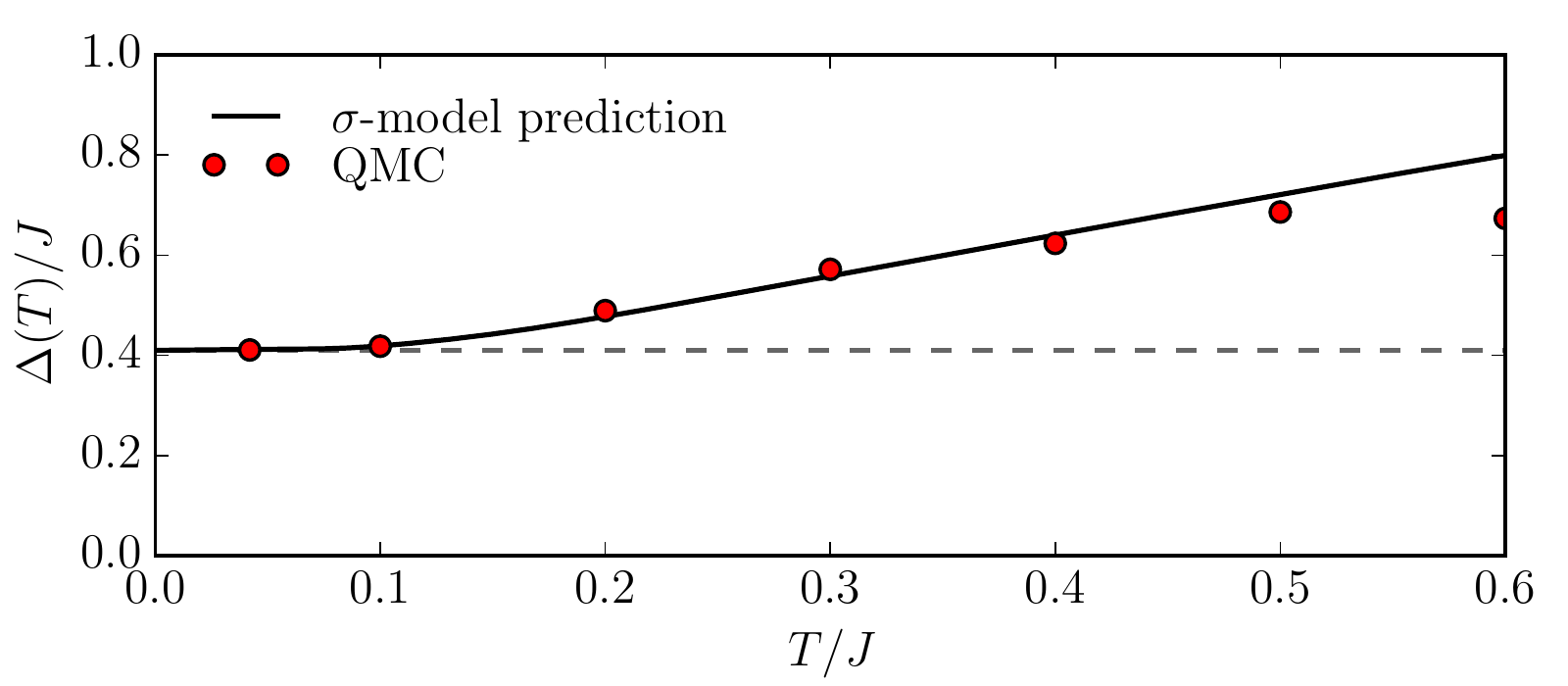}
\caption{Peak energy of the single-magnon peak at $q=\pi$ for finite temperatures, $\Delta(T)$, comparing QMC results  for $L=64$ with PBC (circles) with the $\sigma$-model prediction from Ref.~\onlinecite{Jolicoeur94}  (solid line). }
 \label{fig:band_narrowing} 
\end{figure}

\subsection{Thermal Shift on the Single-Magnon Mode}

In Fig.~\ref{fig:band_narrowing} we compare our results for the energy position of the single-magnon peak in $S(\pi,\omega)$ as obtained from the QMC simulations for $L=64$ with PBC to the results from the $\sigma$-model calculation from Ref.~\onlinecite{Jolicoeur94}.
Like previous work (compare Fig.~3 of Ref.~\onlinecite{Syljuasen08}),
we find that the $\sigma$-model 
describes the numerical data well in the sub-gap temperature region $T/J \lesssim 0.4\,J$, but
we observe deviations at higher temperatures.

\subsection{QMC Results for the Three-Magnon Contribution}

\begin{figure}[h]
\includegraphics[width=0.5\textwidth]{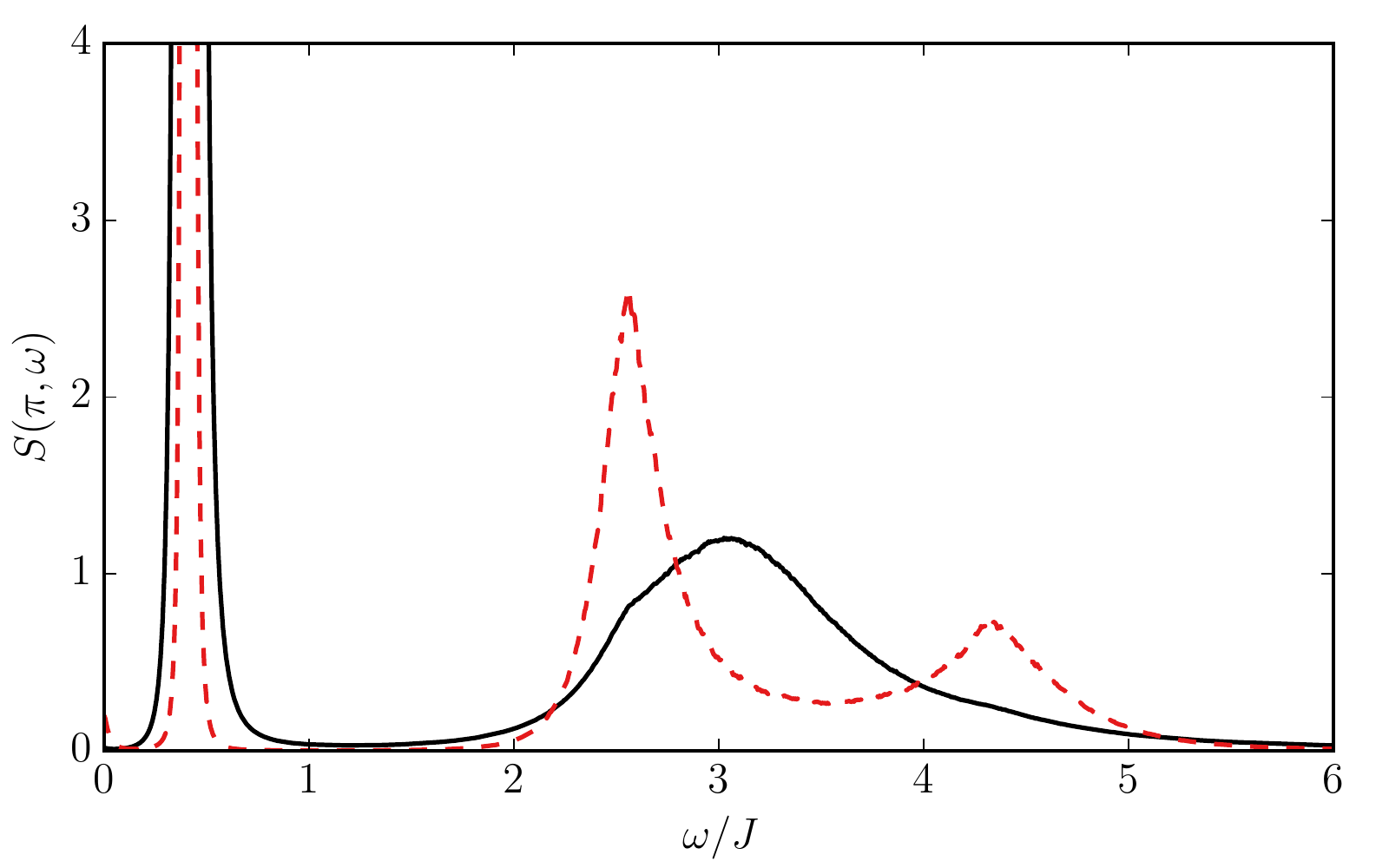}
\caption{Spectral function $S(\pi,\omega)$ at $q=\pi$ as obtained from the analytic continuation of the QMC data for $L=64$ and $T/J=1/24$.  
The black solid line shows   the result  obtained using the  procedure from Ref.~\onlinecite{Beach04}.
When manually forced into the overfitting regime (red dashed line), the analytic continuation yields two distinct peaks.
}
\label{fig:three_magnon_continuum} 
\end{figure}

%
%\vspace{-1cm}
Our QMC results for the three-magnon contribution differ from another
recent QMC investigation \cite{Rahnavard15} and here we would like to
offer an explanation of this discrepancy.
In Fig.~\ref{fig:three_magnon_continuum} we present  the spectral function $S(\pi,\omega)$ as obtained from the analytic continuation of the QMC data for $L=64$ and
 $T/J=1/24$ using the stochastic analytic continuation procedure from Ref.~\onlinecite{Beach04}. Here, we identify, besides the dominant single-magnon peak, a broad single continuum contribution that extends 
 between $\omega/J\approx 1.5$ and $5.5$. This is  in overall accord with the extent and the shape of the three-magnon continuum reported from the zero-temperature DMRG calculations in Ref.~\onlinecite{White08}. In a more recent QMC study of the dynamical spin structure factor of the spin-one chain~\cite{Rahnavard15}, using the maximum entropy approach, the authors obtained a spectral function at $q=\pi$ that exhibits two distinct peaks instead of a single broad continuum. We find that  this result can also
be reproduced based on  our QMC data, if one artificially samples the spectral functions obtained within the stochastic analytic continuation procedure   within  the overfitting region. This suggests that the peculiar two-peak structure seen in the spin spectral function  $S(\pi,\omega)$ in Ref.~\onlinecite{Rahnavard15} may result from an  analytic continuation performed in the overfitting regime.

\end{document}